%% file: bare_jrnl_new_sample4.tex
\documentclass[lettersize,journal]{IEEEtran}
\usepackage{amsmath,amsfonts}
\usepackage{algorithmic}
\usepackage{algorithm}
\usepackage{array}
\usepackage[caption=false,font=normalsize,labelfont=sf,textfont=sf]{subfig}
\usepackage{textcomp}
\usepackage{stfloats}
\usepackage{url}
\usepackage{verbatim}
\usepackage{graphicx}
\usepackage{enumitem}
\usepackage{booktabs}
\usepackage{cite}
\usepackage{pifont}
\usepackage{hyperref}
\usepackage[utf8]{inputenc}
\usepackage{adjustbox}
\begin{document}

\title{Shift-Left High-Level Synthesis Verification via Knowledge-Augmented LLM Agent}

\author{
Zhihan Xiao$^{*}$ \quad 
Hongbing Lang$^{*}$ \quad
Zhe Zhao \quad
Luke Ztz Hu \quad
Songping Mai \ding{41}  
 
\thanks{$^{*}$ Zhihan Xiao and Hongbing Lang contributed equally to this work.}
\thanks{\ding{41} Corresponding Author} 
\thanks{Zhihan Xiao, Zhe Zhao, Luke Ztz Hu, and Songping Mai
are with the Shenzhen International Graduate School, Tsinghua
University, Shenzhen 518055, China (e-mail: xiaozh24@mails.tsinghua.edu.cn;
mai.songping@sz.tsinghua.edu.cn).}
\thanks{Hongbing Lang, Shenzhen Belon Technology Co., Ltd. (e-mail: hblang@belon.cn).} 
}


\maketitle

\begin{abstract}
High-Level Synthesis (HLS) relies on transforming original C specifications into synthesizable HLS-oriented C (HLS-C) implementations. Functional consistency verification between original C specifications and HLS-C implementations is a critical yet labor-intensive task in HLS design flows. While Large Language Models (LLMs) have recently shown promise in automated testbench generation, their stochastic nature often leads to insufficient coverage, inconsistent verification environments, and unreliable equivalence checking results. To address these limitations, we propose a knowledge-augmented, agent-driven shift-left verification framework for automated functional consistency checking between original C and HLS-C implementations before synthesis. The framework introduces a Dual-Tier Consistency Checking mechanism that jointly enforces static structural alignment and dynamic behavioral equivalence between paired testbenches, while integrating symbolic execution and coverage-driven refinement to improve verification completeness. Furthermore, we construct a heterogeneous HLS Verification Knowledge Graph to provide topology-aware reasoning priors for testbench generation, and design an autonomous verification agent to orchestrate iterative refinement and failure diagnosis across heterogeneous toolchains. Experimental results on 107 HLS benchmark pairs demonstrate that the proposed framework achieves 0.9826 average coverage and 0.9533 dynamic consistency, outperforming representative AST-based, retrieval-augmented, and iterative agent-based baselines. 
\href{https://github.com/cz-5f/HLS-LeVeri.git} {https://github.com/cz-5f/HLS-LeVeri.git}
\end{abstract}


\input{sec/intro}
\input{sec/relat}

\input{sec/method}
\input{sec/experiment}

\input{sec/conclusion}

\bibliographystyle{IEEEtran}
\bibliography{main}


\vfill

\end{document}

%% file: sec/intro.tex
\section{Introduction}
\label{sec:introduction}
\IEEEPARstart{W}{ith} the rapid expansion of artificial intelligence, high-performance computing, and embedded systems, hardware acceleration via FPGA and ASIC has become indispensable~\cite{kotari2025review,hu2022survey, zheng2026enhancing}. High-Level Synthesis (HLS) has emerged as a pivotal technology to bridge the software-to-hardware gap, enabling designers to automatically synthesize algorithmic specifications written in C/C++ into Register Transfer Level (RTL) implementations~\cite{zheng2026enhancing}. While HLS significantly accelerates the design space exploration and reduces time-to-market, verifying the correctness of the synthesized hardware remains a manual, time-consuming, and highly error-prone bottleneck~\cite{hlsbasedopt}.

Traditional HLS verification flows heavily rely on C/RTL co-simulation, which validates the post-synthesis RTL against the HLS-C specification~\cite{initiative2015universal, martin2009high, canis2011legup}. However, this paradigm introduces a critical verification gap prior to synthesis. In practice, standard algorithmic implementations (original C) utilize dynamic memory allocation, standard libraries, and floating-point arithmetic, whereas synthesizable code (HLS-C) mandates hardware constraints such as static memory, hardware pragmas (e.g., unrolling, pipelining), customized interfaces (e.g., AXI4-Stream), and arbitrary-precision data types. Because of these structural and semantic discrepancies~\cite{martin2009high}, the logical equivalence between the original C and the HLS-C is frequently compromised. If functional inconsistencies are not isolated early in the design cycle, debugging failures at the cycle-accurate RTL stage becomes prohibitively expensive~\cite{caba2020fpga}. Therefore, shifting left the verification process~\cite{chakraborty2024shift, jain2014early}, namely validating the functional consistency between original C and HLS-C at the source level, is imperative for a reliable HLS design flow.

Recently, Large Language Models (LLMs) have demonstrated immense potential in Electronic Design Automation (EDA), particularly in automated code refactoring and testbench generation~\cite{hlstrans, xu2025hlstester, hls-eval}. Benefiting from their strong semantic understanding and code synthesis capabilities, LLMs can substantially reduce the manual effort required for verification environment construction, interface adaptation, and stimulus generation, making them highly attractive for shift-left HLS verification. However, directly applying pure LLM-based generators to bit-exact hardware verification still reveals fundamental limitations~\cite{hls-eval,lu2024rtllm}. Since LLMs are inherently stochastic, they frequently hallucinate and lack the deterministic mathematical reasoning required to resolve complex algebraic path constraints~\cite{banerjee2025llms}. Consequently, zero-shot generated testbenches often saturate at low coverage plateaus, failing to trigger deep algorithmic corner cases or adequately explore complex control-data interactions~\cite{xu2025hlstester,herklotz2021formal,chouksey2020verification}. 
More critically, existing LLM-driven approaches largely treat these objectives independently, lacking a unified mechanism that integrates deterministic reasoning, coverage-guided refinement, and semantic alignment enforcement within a closed verification loop~\cite{hlspilot, Graph-RAG, KnowledgeGraphBased, synthai, ConfiBench}.

\begin{figure}[t]
  \centering
  \includegraphics[width=0.98\linewidth]{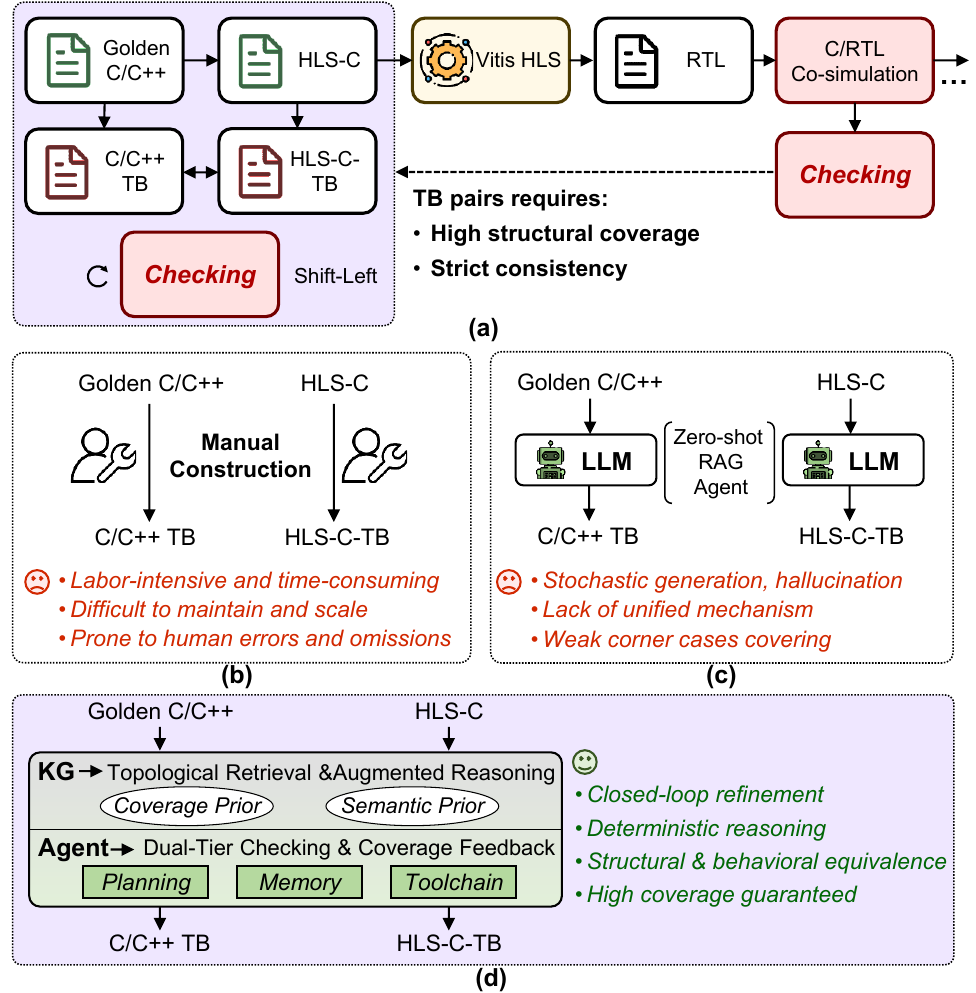}
  \caption{(a) Shift-Left Verification Process. (b) Manual Testbench Construction. (c) Existing LLM-based Testbench Generation. (d) Our Proposed Knowledge-Augmented LLM Agent.
  }
  \label{fig:intro}
\end{figure}

As shown in Fig.~\ref{fig:intro}, unlike previous methods, we propose an automated, closed-loop, shift-left consistency verification framework that systematically validates the functional equivalence between a software C specification and its HLS-oriented counterpart prior to synthesis. Due to the substantial structural and semantic discrepancies between original C and HLS-C implementations, independently generated testbenches may introduce hidden asymmetries in input stimuli, control-flow logic, or data dependencies~\cite{martin2009high,canis2011legup}. Therefore, an effective shift-left verification framework must simultaneously address two tightly coupled objectives: \textit{(i)} achieving sufficiently high structural coverage to expose deep corner-case behaviors, and \textit{(ii)} enforcing strict consistency between the verification environments themselves. 

Recognizing that program equivalence checking is only mathematically sound if the verification environments themselves are strictly aligned, our framework introduces a Dual-Tier Consistency Checking mechanism. This mechanism first guarantees the static structural isomorphism of the generated testbenches by analyzing input stimuli, Control Flow Graph (CFG), and Data Dependency Graph (DDG) before executing a cycle-accurate dynamic behavioral co-simulation. To orchestrate this, we design an autonomous verification Agent that moves beyond conventional one-shot prompting. To eliminate LLM hallucinations and escape the coverage plateaus typical of stochastic generation, the Agent tightly integrates the LLM with KLEE~\cite{KLEE} (a symbolic execution engine) for deterministic path constraint solving and \textit{gcov} for dynamic structural coverage feedback. Furthermore, to guide the LLM's initial scaffolding and subsequent iterative repairs, we introduce a novel HLS Verification Knowledge Graph (KG). By retrieving verified structural invariants and semantic alignment rules from a purpose-built HLS dataset, the Knowledge Graph provides structural priors that strictly enforce cross-domain testbench consistency, transforming isolated code generation into a robust, self-refining hardware verification pipeline. Extensive evaluations demonstrate that our framework outperforms state-of-the-art Retrieval-Augmented Generation (RAG) and iterative agent paradigms.

The main contributions are summarized below:
\begin{itemize}
    \item We propose a Dual-Tier (static and dynamic) consistency checking methodology that decouples testbench generation artifacts from genuine hardware design flaws, ensuring rigorous C vs. HLS-C equivalence.
    \item We design an Verification Agent that forms a closed-loop system. By integrating symbolic execution and execution profiling, the agent deterministically augments stimuli to achieve high-coverage verification sign-off.
    \item We construct a novel, heterogeneous HLS Verification Knowledge Graph that provides structural priors to the LLM, enforcing cross-domain testbench consistency without relying on superficial code similarity.
\end{itemize}

The following sections elaborate the design details and experimental validation of this work. Section \ref{sec:background} introduces the fundamental background of high-level synthesis verification. Section \ref{sec:method} presents the full methodology of the proposed framework. Performance evaluation is presented in Section \ref{sec:experiment}. Finally, Section \ref{sec:conclusion} concludes this article and discusses potential future research directions.

%% file: sec/relat.tex
\section{Background and Related Work}
\label{sec:background}

\subsection{Existing HLS Verification Methods}
While commercial tools like Vitis HLS provide robust C/RTL co-simulation, this assumes the HLS-C code is already functionally flawless. In academic research, Karfa \textit{et al.}~\cite{karfa2006formal} proposed a systematic formal verification framework combining model checking and theorem proving to validate the entire HLS flow from algorithmic descriptions to RTL implementations. Peñalba \textit{et al.}~\cite{mendias2002study} presented an efficient formal HLS verification methodology that reconstructs the synthesis process within a formal synthesis framework according to reports generated by external HLS tools, achieving verification overheads of only around 5\%. In addition, Xu \textit{et al.}~\cite{xu2025hlstester} proposed the HLSTester framework, which introduces execution-profile monitoring to track the behaviors of critical variables and prove the semantic equivalence between two C programs through symbolic execution and static analysis.

The integration of LLMs into the EDA toolchain has seen significant traction. Recent frameworks, such as C2HLS ~\cite{c2hlsc} and HLSPilot~\cite{hlspilot}, leverage prompt engineering and intermediate representations to automatically translate C/C++ to HLS-C. In the domain of functional verification, frameworks like ConfiBench~\cite{ConfiBench} have pioneered the use of LLMs for automated HDL testbench generation, employing execution feedback to iteratively refine the test stimuli. Wang \textit{et al.} ~\cite{wang2025hlsdebugger} proposed the HLSDebugger, constructing 3M labeled HLS error examples to train the model, which is used for automatically locating and fixing logical errors in HLS-Code.

Existing LLM-driven generation methods primarily focus on syntactic correctness and synthesis pass rates, often neglecting bit-exact functional equivalence. Additionally, due to the non-deterministic nature of LLMs, the generated testbenches for the original C and HLS-C may possess asymmetric stimulus logic, leading to false positives during verification.


\subsection{Symbolic Execution and Coverage-Driven Verification}
In hardware-oriented verification scenarios, structural coverage metrics, including statement coverage, branch coverage, and call coverage, are essential indicators for evaluating verification completeness~\cite{zhu1997software}. Existing LLM-driven test generation methods~\cite{ConfiBench, wang2025hlsdebugger, zhang2025llm4dv} primarily rely on randomized or heuristic stimulus synthesis, often failing to activate rare execution branches or deeply nested control flows. Symbolic execution alleviates this limitation by transforming coverage exploration into a constraint-solving problem. 

Formally, symbolic execution models a program state as a tuple: $\sigma = \langle \ell, \mu, PC_{\pi} \rangle$, where $\ell$ denotes the current program counter, $\mu$ represents the symbolic memory store, and $PC_{\pi}$ is the accumulated path constraint over symbolic inputs. 
During execution, branch instructions fork the symbolic state into multiple feasible execution paths, while satisfiability modulo theories (SMT)~\cite{barrett2018satisfiability} solvers are employed to determine whether the associated constraints are satisfiable. If satisfiable, the solver derives concrete assignments that satisfy path constraints and drive execution toward unexplored states.

In this field, Cadar \textit{et al.}~\cite{KLEE} proposed KLEE, one of the most widely adopted symbolic execution engines, which systematically explores program paths to generate high-coverage test cases. Sen \textit{et al.}~\cite{sen2005cute} further introduced concolic testing, which combines concrete execution with symbolic reasoning to improve scalability and path exploration efficiency. While tools like HLSTester~\cite{xu2025hlstester} have utilized KLEE to compare C and HLS-C behaviors, they lack an autonomous closed-loop refinement mechanism.

\subsection{Knowledge-Augmented Program Analysis}

To mitigate the hallucination problem in LLMs, RAG has been widely adopted in code generation tasks~\cite{arks,pkg_rag,kg2rag}. 
Existing approaches commonly retrieve semantically related code fragments from external repositories using vector similarity or Abstract Syntax Tree (AST) retrieval mechanisms. 
Recent studies further demonstrate that graph-structured program representations can substantially improve retrieval precision by modeling higher-order structural dependencies beyond lexical similarity~\cite{pkg_rag,cast_rag}. 

In program analysis, Knowledge Graph (KG) are typically constructed from syntactic and semantic program entities, including functions, variables, control-flow structures, and compiler directives. 
Such graphs are commonly represented as collections of relational triplets: $\mathcal{G} = \langle \mathcal{V}, \mathcal{E}, \mathcal{R} \rangle$, where $\mathcal{V}$ denotes the set of graph entities, $\mathcal{E}$ represents the edges among entities, and $\mathcal{R}$ denotes the associated relation types. 
Compared with flat text retrieval, graph-structured representations preserve topological dependencies and enable multi-hop reasoning across heterogeneous program components~\cite{kg2rag}. 


%% file: sec/method.tex
\section{Methodology}
\label{sec:method}

\subsection{Architecture Overview}
\label{sec:overview}

Given a pair of programs, namely a original C/C++ specification $P_c$ and its HLS-oriented counterpart $P_h$, the objective of this work is to automatically construct a high-coverage and semantically consistent testbench pair $(TB_c, TB_h)$, such that functional equivalence between $P_c$ and $P_h$ can be rigorously validated. Rather than modifying or repairing the source code, the proposed framework focuses exclusively on verification, aiming to provide reliable evidence of equivalence or to expose concrete counterexamples when discrepancies exist.

\begin{figure*}[t]
    \centering
    \includegraphics[width=\linewidth, draft=false]{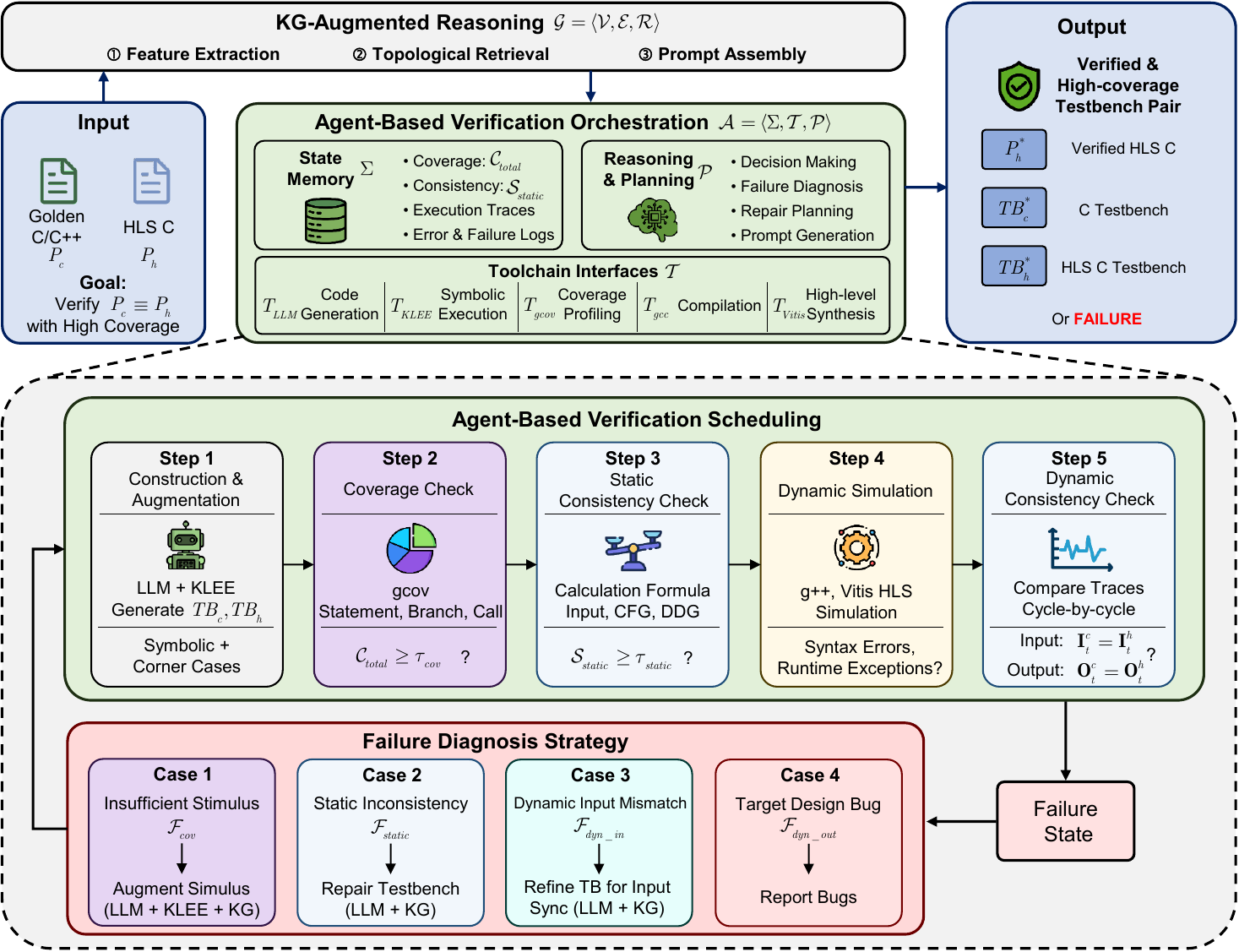}
    \caption{Overview of the proposed KG-augmented, agent-based verification framework for functional consistency checking between original C and HLS-C implementations. The framework integrates a coverage-driven verification scheduling strategy with a closed-loop orchestration mechanism.
}
    \label{fig:agent_workflow}
\end{figure*}

As illustrated in Fig.~\ref{fig:agent_workflow}, we formulate this task as a knowledge-augmented, agent-driven verification pipeline. The workflow integrates 
a Dual-Tier Consistency Checking mechanism and a Coverage-driven Testbench Construction module. On this basis, an autonomous agent orchestrates the process in a closed loop, leveraging coverage metrics, trace comparisons, and Knowledge Graph priors to iteratively refine the testbench until coverage and consistency criteria are satisfied or the maximum iteration budget is reached.

\subsection{Functional Consistency Verification Objective}
\label{sec:consistency}
\subsubsection{Problem Definition}
Formally, complete functional equivalence requires that for any given input vector $\mathbf{x}$ drawn from the entire valid input space $\mathcal{X}$, the outputs of both programs are strictly identical:
\begin{equation}
\forall \mathbf{x} \in \mathcal{X}, \mathcal{M}_c(P_c, \mathbf{x}) \equiv \mathcal{M}_h(P_h, \mathbf{x}),
\end{equation} 
where $\mathcal{M}_c$ and $\mathcal{M}_h$ represent the execution semantics of the respective programs. 
We approximate the formal proof through a rigorous, bounded test-based verification methodology.
The fundamental challenge, however, arises from the reliance on LLMs to generate these testbenches. Since LLMs are non-deterministic, we cannot inherently trust that $TB_c$ and $TB_h$ are semantically identical.
Thus, we formalize a Dual-Tier Consistency Checking mechanism.

\subsubsection{Dual-Tier Consistency Check} Our methodology decomposes consistency checking into two sequential phases: a static structural precondition check and a dynamic behavioral equivalence check. 

\textit{(i) Testbench Consistency (Static Precondition):} Before invoking cycle-accurate simulation, we must ensure that the generated testbenches are functionally isomorphic. We evaluate static consistency across three dimensions by analyzing the AST of $TB_c$ and $TB_h$.

\noindent \ding{172} Input Stimulus Consistency ($\mathcal{S}_{IO}$): We extract the sub-trees related to data injection (e.g., variable declarations, literals, and array initializations). To measure the syntactic alignment of the input generation, we apply the normalized Levenshtein distance $\text{Dist}_{Lev}(\cdot)$  over the linearized AST node sequences $S_c$ and $S_h$: 
\begin{equation}
\mathcal{S}_{IO} = 1 - \frac{\text{Dist}_{Lev}(S_c, S_h)}{\max(|S_c|, |S_h|)},
\end{equation}
where $|S_c|$ and $|S_h|$ denote the lengths of the respective linearized node sequences.

\noindent \ding{173} CFG Consistency ($\mathcal{S}_{CFG}$): To ensure the testbenches exercise the identical branching logic, we compute a weighted similarity of their CFGs. Let $\text{Sim}_{ctrl}$ be the distribution similarity of control nodes (e.g., IfStmt, ForStmt) and $\text{Sim}_{nest}$ be the normalized difference in maximum nesting depth. The CFG consistency is defined as:
\begin{equation}
\mathcal{S}_{CFG} = w_{ctrl} \cdot \text{Sim}_{ctrl} + w_{nest} \cdot \text{Sim}_{nest},
\end{equation}
where $w_{ctrl}$ and $w_{nest}$ are pre-defined weights that balance the importance of control distribution and nesting hierarchy.

\noindent \ding{174} DDG Consistency ($\mathcal{S}_{DDG}$): To capture semantic equivalence, we perform a Reaching Definition analysis to extract the sets of Def-Use chains, denoted as $E_c$ and $E_h$. The DDG similarity is calculated using the intersection of structural data-flow edges:
\begin{equation}
\mathcal{S}_{DDG} = \frac{|E_c \cap E_h|}{|E_c \cup E_h|},
\end{equation}
where $|\cdot|$ denotes the cardinality of the edge sets.

The overall Static Consistency Score ($\mathcal{S}_{static}$) is aggregated via a weighted linear combination. A testbench pair is strictly rejected if $\mathcal{S}_{static}$ falls below a rigorous threshold $\tau_{static}$, preventing invalid simulations.
\begin{equation}
\mathcal{S}_{static} = \omega_1 \mathcal{S}_{IO} + \omega_2 \mathcal{S}_{CFG} + \omega_3 \mathcal{S}_{DDG} \geq \tau_{static},
\end{equation}
where $\omega_1, \omega_2, \omega_3$ are empirically determined weights.

\textit{(ii) Behavioral Consistency (Runtime Equivalence):} Once the testbench pair satisfies the static precondition, we execute a cycle-by-cycle dynamic co-simulation.
Each execution produces a time-ordered trace $Traces=\{(\mathbf{I}_t, \mathbf{O}_t)\}_{t=0}^{t_{max}}$, capturing the input-output behavior at each cycle. Let $\mathbf{I}^c_t$ and $\mathbf{I}^h_t$ be the exact input signals injected at cycle $t$, and $\mathbf{O}^c_t$ and $\mathbf{O}^h_t$ denote the corresponding outputs recorded in the execution traces $Traces_c$ and $Traces_h$.  We define a strict two-step runtime assertion protocol:

\noindent \ding{172} Input Equivalence Check: $\forall t, \mathbf{I}^c_t \equiv \mathbf{I}^h_t$. If this condition is violated during runtime despite passing the static checks, the testbench pair contains hidden semantic divergences. We classify this as a testbench bug, triggering a feedback loop to the LLM agent to rectify the testbench generation.

\noindent \ding{173} Output Equivalence Check: Given condition \ding{172} holds, $\forall t, \mathbf{O}^c_t \equiv \mathbf{O}^h_t$. If the inputs are strictly aligned but the outputs diverge at any cycle $t$, we have deterministically isolated a design bug between the original C and the HLS-C source code. 

By enforcing this Dual-Tier Consistency Checking mechanism, our framework cleanly decouples verification environment errors from actual HLS translation errors, ensuring a highly reliable automated shift-left verification flow.

\subsection{Coverage-Driven Testbench Construction}
\label{sec:coverage}
Our framework introduces a closed-loop, symbolic execution-guided construction architecture that systematically orchestrates testbench generation, ensuring high-fidelity stimulus injection and maximum structural coverage.


\subsubsection{Symbolic Execution-Guided Stimulus Augmentation}
To overcome the coverage ceiling of stochastic generation, we integrate KLEE, a robust symbolic execution engine, to systematically augment the verification stimulus. In this architecture, the LLM provides the syntactical prior and the test environment, while KLEE guarantees path completeness and specific numerical injections.

Rather than executing $P_c$ with concrete integer or floating-point values, KLEE treats the input parameters as symbolic variables. As it traverses the CFG of the program, KLEE constructs a set of path constraints $PC_{\pi}$ for each feasible execution path $\pi \in \Pi$, where $\Pi$ denotes the set of all feasible paths. At each branch instruction, the execution state forks, and the corresponding constraints are updated as follows:
\begin{equation}
PC_{\pi_{\text{true}}} = PC_{\pi} \wedge \text{cond}, \quad
PC_{\pi_{\text{false}}} = PC_{\pi} \wedge \neg \text{cond},
\end{equation}
where $\text{cond}$ denotes the branch predicate.


\subsubsection{Closed-Loop Coverage Orchestration}
The LLM and symbolic execution engines are bound together by an automated, dynamic profiling loop utilizing \textit{gcov}.
We formalize our verification completeness objective using a composite coverage metric ($\mathcal{C}_{total}$), evaluating three critical dimensions of the program's topology:
\begin{enumerate}[label=\textit{(\roman*)}]
    \item Statement Coverage ($\mathcal{C}_{stmt}$): The ratio of executable lines successfully triggered by the testbench.
    \item Branch Coverage ($\mathcal{C}_{branch}$): The ratio of evaluated control-flow edges (both true and false outcomes).
    \item Call Coverage ($\mathcal{C}_{call}$): The ratio of internal sub-functions and module instantiations successfully invoked.
\end{enumerate}
The overall coverage score is evaluated as a weighted sum:
\begin{equation}
    \mathcal{C}_{total} = \omega_{stmt} \mathcal{C}_{stmt} + \omega_{branch} \mathcal{C}_{branch} + \omega_{call} \mathcal{C}_{call}.
\end{equation} 

If $\mathcal{C}_{total} < \tau_{cov}$, the framework extracts uncovered statements, branches, and function calls identified by \textit{gcov}. These coverage deficits are converted into targeted prompts, enabling the LLM to augment the existing testbench with additional stimuli and KLEE-derived vectors. The loop iterates until $\mathcal{C}_{total} \geq \tau_{cov}$ or a predefined timeout is reached. 

\subsection{Agent-Based Verification Orchestration}
\label{sec:agent}
While the mathematical bounds of consistency (Section ~\ref{sec:consistency}) and the mechanisms for stimulus augmentation (Section~\ref{sec:coverage}) form the theoretical foundation of our framework, executing this heterogeneous pipeline manually is highly error-prone and practically infeasible. A complete verification sign-off requires seamless data passing and semantic alignment across disjoint toolchains, including LLMs, formal constraint solvers (KLEE), coverage profilers (\textit{gcov}), standard compilers (g++), and hardware synthesizers (Vitis HLS). To unify these disparate processes, we design an autonomous verification Agent, acting as a stateful orchestrator.

\subsubsection{Stateful Orchestration and Tool Invocation}
As illustrated in Fig. \ref{fig:agent_workflow}, we formally define the Agent as a tuple $\mathcal{A} = \langle \Sigma, \mathcal{T}, \mathcal{P} \rangle$. 
$\Sigma$ represents the Agent's dynamic state memory, which continuously tracks the current coverage metrics $\mathcal{C}_{total}$, consistency scores $\mathcal{S}_{static}$, execution traces, and historical error logs. $\mathcal{T}$ represents the integrated toolchain interfaces $\{T_{LLM}, T_{KLEE}, T_{gcov}, T_{gcc}, T_{Vitis}\}$. The Agent autonomously invokes these tools, parses their raw outputs (e.g., standard error streams, AST JSONs, profiling data), and abstracts them into an internal semantic representation. $\mathcal{P}$ is the reasoning and planning engine driven by the LLM, responsible for deciding the next state transition based on the current verification status.

\subsubsection{Agent-Based Verification Scheduling}
Formally, each iteration follows a fixed sequence: \textit{(i)} coverage evaluation and stimulus augmentation, \textit{(ii)} static consistency verification, and \textit{(iii)} dynamic co-simulation and equivalence checking. Any failure detected in later stages triggers a refinement loop that returns to the coverage evaluation stage, ensuring progressive improvement toward both coverage completeness and functional correctness.

\subsubsection{Autonomous Failure Diagnosis Strategy}
Our Agent employs a deterministic Failure Diagnosis Strategy, classifying anomalies into a strict debug taxonomy to execute targeted, multi-step refinements. Let $\mathcal{F}$ denote a failure state detected during the orchestration loop. The Agent evaluates $\mathcal{F}$ against four distinct topological cases to dictate its iterative repair strategy:

\textbf{Case 1: Insufficient Stimulus ($\mathcal{F}_{cov}$).} \textit{Condition:} The coverage profiling indicates $\mathcal{C}_{total} < \tau_{cov}$. \textit{Agent Action:} The Agent identifies this not as a code bug, but as a testbench deficiency. It extracts the unreached branch conditions from $T_{gcov}$, feeds them into $T_{KLEE}$ for path constraint resolution, and prompts the LLM to structurally append the newly derived corner-case vectors into $TB_c$ and $TB_h$.

\textbf{Case 2: Static Inconsistency ($\mathcal{F}_{static}$).} \textit{Condition:} The pre-simulation static consistency score $\mathcal{S}_{static} < \tau_{static}$. \textit{Agent Action:} The Agent intercepts the pipeline before simulation, avoiding wasted compute cycles.
The LLM is prompted to strictly realign the anomalous testbench structure to match its counterpart.

\textbf{Case 3: Dynamic Input Mismatch ($\mathcal{F}_{dyn\_in}$).} \textit{Condition:} During cycle-by-cycle co-simulation, the input sequences diverge ($\exists t, \mathbf{I}^c_t \neq \mathbf{I}^h_t$). \textit{Agent Action:} 
The Agent masks the target design $P_h$, feeds the diverging input trace back to the LLM, and refines the testbench generation logic to enforce bit-exact input synchronization.

\textbf{Case 4: Target Design Bug ($\mathcal{F}_{dyn\_out}$).} \textit{Condition:} Input stimuli are perfectly synchronized ($\forall t, \mathbf{I}^c_t \equiv \mathbf{I}^h_t$), but the output traces diverge ($\exists t, \mathbf{O}^c_t \neq \mathbf{O}^h_t$). \textit{Agent Action:} The Agent deterministically isolates a functional discrepancy between $P_c$ and $P_h$. It reports the failing cycle $t$ and the corresponding counterexample $\mathbf{x}_{fail}$ to assist downstream debugging or repair processes.

Additionally, if compilation or simulation fails (e.g., syntax errors, runtime exceptions), the Agent classifies the failure as an execution error and triggers a repair loop by analyzing runtime logs and prompting the LLM to fix the corresponding issues in the testbench or source code.

\begin{figure*}[t]
    \centering
    \includegraphics[width=\linewidth, draft=false]{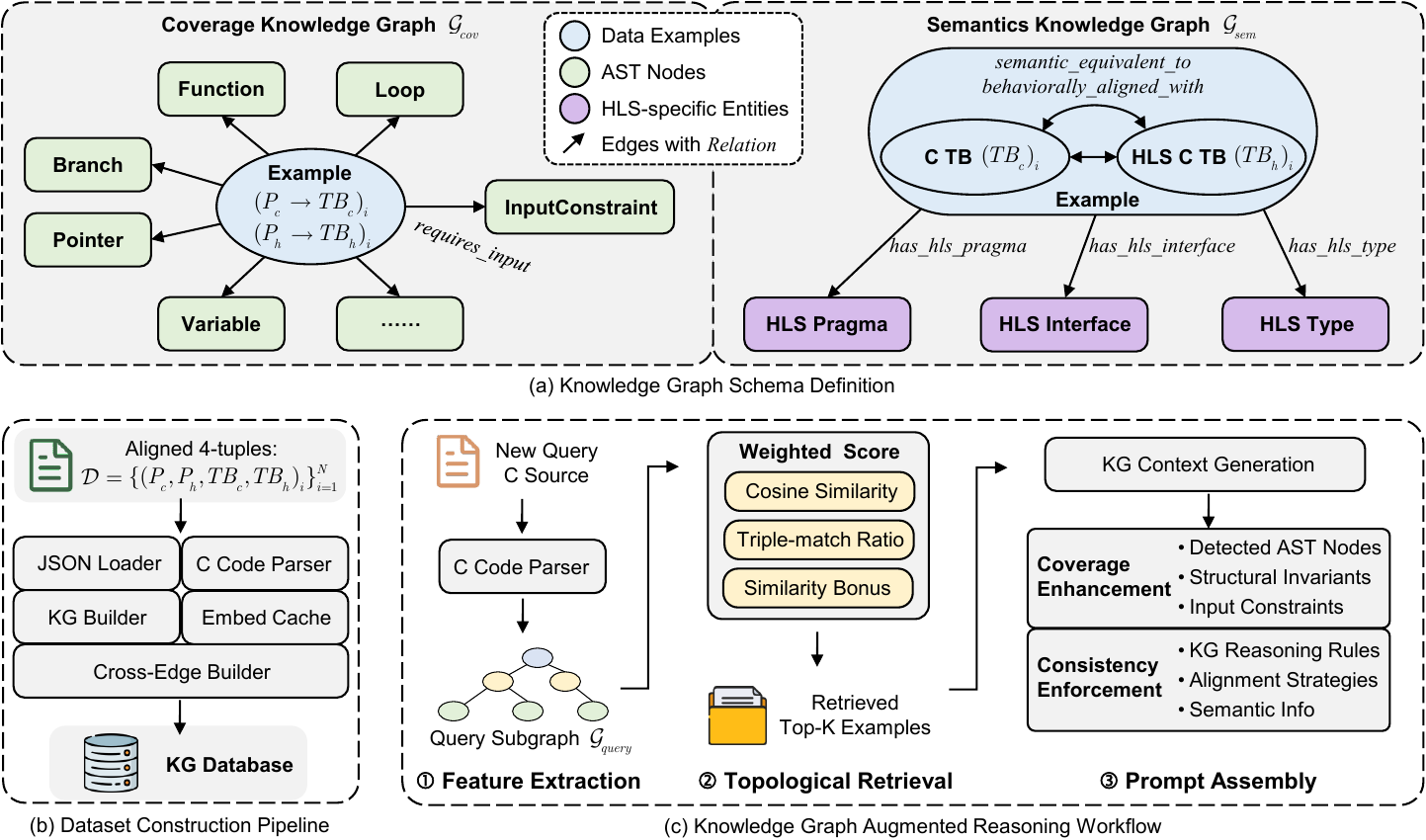}
    \caption{Overview of the proposed KG-augmented verification framework. During inference, graph-retrieved reasoning priors guide coverage-driven stimulus generation and enforce cross-domain testbench consistency.
}
    \label{fig:kg_schema}
\end{figure*}

\subsection{Knowledge Graph Augmented Verification}
\label{sec:kg}
Post-generation feedback incurs significant computational overhead. In a pure trial-and-error paradigm, the Agent lacks a priori knowledge of common verification patterns. To enhance the first-pass construction quality, we augment the Agent with an experience reuse module powered by a novel HLS Verification Knowledge Graph. 

\subsubsection{Dataset Construction Pipeline}
We systematically construct a domain-specific corpus based on the Chrysalis-HLS~\cite{chrysalis} dataset. The original Chrysalis-HLS dataset provides approximately 1,000 foundational hardware-accelerated code modules, and it contains the ground-truth HLS-C implementations. To further improve structural diversity, we apply data augmentation techniques, including parameterized array dimensions and varying loop unrolling factors. For each HLS-C module, the Agent constructs a candidate pure C reference model ($P_c$), which is subsequently validated through the proposed consistency framework. Subsequently, the Agent utilizes the exhaustive KLEE and \textit{gcov} loop to generate highly optimized, perfectly aligned testbench pairs ($TB_c, TB_h$). 

Finally, to ensure high-quality supervision, we perform strict filtering based on the previously defined Dual-Tier Consistency metrics, enforcing both functional equivalence and coverage adequacy. After this verification-driven pruning, 1,597 high-confidence and strictly aligned data 4-tuples are retained, yielding the final dataset: $
    \mathcal{D} = \{(P_c, P_h, TB_c, TB_h)_i\}_{i=1}^N$.


Overall, the dataset exhibits a balanced coverage across compute-intensive kernels (e.g., machine learning and numerical computation) and system-level operators (e.g., streaming, control logic, and memory subsystems), ensuring that the proposed framework is evaluated over a broad spectrum of realistic HLS design patterns.

\subsubsection{Knowledge Graph Schema Definition}

Instead of storing raw code text, which leads to inefficient semantic retrieval, the dataset $\mathcal{D}$ is abstracted into a heterogeneous directed Knowledge Graph, formally defined as $\mathcal{G} = \langle \mathcal{V}, \mathcal{E}, \mathcal{R} \rangle$. Here, $\mathcal{V}$ denotes the set of syntactic and semantic entity nodes (e.g., Function, Loop, Pointer, HLSPragma), and $\mathcal{E}$ denotes the set of directed edges associated with relation types $\mathcal{R}$ (e.g., \textit{requires\_input}, \textit{semantic\_equivalent\_to}, \textit{has\_hls\_pragma}).

As illustrated in Fig.~\ref{fig:kg_schema}, the graph is logically partitioned into two functionally distinct subgraphs:

\textit{(i) Coverage Knowledge Graph} ($\mathcal{G}_{cov}$): This subgraph captures the relationship between program structural patterns 
and the corresponding stimulus strategies required to achieve high coverage. It encodes historical path constraints and corner-case input distributions derived from symbolic execution.

\textit{(ii) Semantics Knowledge Graph} ($\mathcal{G}_{sem}$): This subgraph models the structural and behavioral alignment rules between paired testbenches ($TB_c$, $TB_h$) within the same instance. It encodes interface-level and data representation consistency constraints, such as type alignment, data formatting, and synchronization protocols.
Formally, for a given instance $i$, $\mathcal{G}_{sem}$ encodes the verified alignment relationship between the paired testbenches $(TB_c^{(i)}, TB_h^{(i)})$.

Specifically, the structural patterns of stimulus generation are isomorphic:
\begin{equation}
\Phi(TB_c^{(i)}) \cong \Phi(TB_h^{(i)}),
\end{equation}
where $\Phi(\cdot)$ extracts the input-related AST substructures.


\subsubsection{KG-Augmented Reasoning with Structural Priors}
The proposed approach extracts abstract reasoning patterns and structural invariants from the Knowledge Graph. Given a new program pair $(P_c, P_h)$, the Agent performs the following steps:

\textit{(i) Feature Extraction:} The system parses the input programs and constructs a localized query subgraph $\mathcal{G}_{query}$, capturing key interface signatures, data types, and control-flow characteristics.

\textit{(ii) Topological Retrieval:} A graph traversal is performed to identify subgraphs in $\mathcal{G}_{cov}$ and $\mathcal{G}_{sem}$ that exhibit high structural similarity to $\mathcal{G}_{query}$. For a given query graph $\mathcal{G}_{query}$ and a candidate example node $v \in \mathcal{V}$, the retrieval score $S(query, v)$ is formulated as a weighted sum of semantic, topological, and neighborhood bonus features:
\begin{equation}
\begin{split}
    S(query, v) = {} & w_{emb} \cdot Sim_{emb}(query, v) \\
                     & + w_{topo} \cdot Sim_{topo}(query, v) \\
                     & + w_{sem} \cdot Bonus_{sem}(v),
\end{split}
\end{equation}
where $Sim_{emb}$ represents the deep feature cosine similarity, and $Sim_{topo}$ is the normalized triple-match ratio defined as the intersection of identical AST structural edges between the query and candidate $v$. To encourage the retrieval of highly reliable verification templates, $Bonus_{sem}$ propagates fractional similarity scores from the candidate's semantically equivalent neighbors. 

\textit{(iii) Reasoning Prompt Assembly:} Instead of injecting raw code, the retrieved subgraphs are abstracted into deterministic reasoning rules that guide the LLM during testbench generation. This mechanism serves as a structural prior for both verification objectives, i.e., coverage enhancement and testbench consistency enforcement.


By incorporating graph-derived reasoning priors, the framework significantly reduces the search space of valid testbenches. The Knowledge Graph provides high-level structural guidance, while the Agent ensure fine-grained correctness and coverage, resulting in a synergistic and efficient verification pipeline.

%% file: sec/experiment.tex
\section{Experimental Evaluation}
\label{sec:experiment}
\subsection{Experimental Setup}
\label{sec:exp_setup}

\subsubsection{Implementation Details}
We implement the proposed framework as an end-to-end automated verification pipeline targeting shift-left HLS verification, where functional correctness is validated prior to synthesis. For hardware-oriented validation, we adopt Vitis HLS 2019.1 as the backend to compile and simulate HLS-C designs, enabling cycle-accurate co-simulation and trace extraction. 

During the offline dataset construction phase, the Agent operates exhaustively without Knowledge Graph priors, with the maximum number of iterative refinement rounds set to 6. During the online inference phase (i.e., the experimental evaluation), we strictly limit the maximum iterations to 3 rounds to evaluate the efficiency of the KG-augmented reasoning. For non-iterative or zero-shot baseline methods, we adopt a \textit{\textbf{pass@3}} evaluation protocol to align the computational budget with our 3-round iterative Agent. Furthermore, the evaluation thresholds are uniformly strictly maintained across both dataset construction and inference phases: the coverage threshold is set to $\tau_{cov} = 0.95$, and the static consistency threshold is set to $\tau_{static} = 0.75$, where both metrics are normalized to the range $[0,1]$.

\subsubsection{Baseline Methods}
We compare against representative methods from three categories of prior work in testbench generation and HLS-related code intelligence.
\textit{(i)} For AST-based Knowledge Graph Methods, we compare our framework with the AST-based Graph-RAG approach proposed by Chinthareddy~\cite{Graph-RAG}. 
\textit{(ii)} Retrieval-Augmented Generation Methods: SynthAI~\cite{synthai}, which retrieve relevant code snippets or examples to guide LLM generation.
\textit{(iii)} Iterative Agent-based Method: ConfiBench~\cite{ConfiBench}, which refine generated testbenches through execution feedback.

For all baselines, we re-implement or adapt their pipelines to the HLS verification setting to the best extent possible. When full implementations are not publicly available, we faithfully reproduce their core design principles based on published descriptions to ensure a fair comparison.

\subsubsection{Evaluation Metrics}
We evaluate all methods using four complementary metrics that reflect correctness, coverage, and consistency of the generated testbenches.

\textit{(i) Syntax Correctness.}
We measure the ratio of testbench pairs that can be successfully compiled and executed without syntax or runtime errors:
$
\text{Syntax Rate} = N_{\text{valid}}/{N_{\text{total}}}
$.

\textit{(ii) Coverage Score.}
Following the formal definition in Section~\ref{sec:coverage}, we evaluate the testbench stimulus quality using the composite coverage metric $\mathcal{C}_{total}$, which aggregates statement, branch, and function call coverage based on the previously established weighting scheme.

\textit{(iii) Static Consistency Score.}
We evaluate the structural alignment between generated testbench pairs using the static consistency metrics $\mathcal{S}_{IO}$, $\mathcal{S}_{CFG}$, and $\mathcal{S}_{DDG}$ defined in Section~\ref{sec:consistency}.

\textit{(iv) Dynamic Consistency Rate.}
We measure whether the generated testbench pair produces bit-exact identical execution traces under co-simulation: $
\text{Dynamic Rate} = N_{\text{match}} / N_{\text{total}}$, 
where a testbench pair is counted as valid only if $\forall t, \mathbf{I}^c_t \equiv \mathbf{I}^h_t$. 
We emphasize that this metric evaluates testbench-level functional consistency without inspecting the source. 

\subsubsection{Benchmarks}
We evaluate our framework on the HLSTrans~\cite {hlstrans} benchmark, which provides comprehensive paired C and HLS-C programs collected from multiple established HLS datasets or benchmarks, including HLSFactory~\cite{hlsfactoryc}, ForgeBench~\cite{forgebench}, C2HLS~\cite{c2hlsc}, MachSuite~\cite{machsuite}, PP4FPGA~\cite{pp4fpga}, etc. 

We perform a strict filtering process to extract a subset of functionally independent program pairs. Specifically, we remove duplicated patterns and parameterized variants, and retain 107 distinct program pairs that serve as \textit{verification targets}. These benchmarks cover a wide spectrum of HLS kernels, including numerical computation, signal processing, and control-intensive designs. Each pair $(P_c, P_h)$ is treated as a test instance without assuming prior knowledge of functional equivalence, aligning with our goal of validating testbench correctness and detecting potential inconsistencies in a shift-left verification setting. 

\input{mic/ablation1}
\input{mic/datascore}

\subsection{Dataset Quality Evaluation}
\label{sec:Dataset}
To validate the quality and completeness of the constructed dataset $\mathcal{D}$, we perform both structural comparison against existing HLS-related datasets and quantitative evaluation based on verification-oriented metrics.

Table~\ref{tab:dataset_quality} (a) compares our dataset with representative HLS datasets and benchmarks, including HLSDataset~\cite{hlsdataset}, HLS-Eval\cite{hls-eval}, HLSTrans~\cite{hlstrans}, HLSPilot~\cite{hlspilot}. Existing datasets typically provide only partial components of the verification pipeline.
In contrast, our dataset is the first to provide a complete 4-tuple structure $(P_c, P_h, TB_c, TB_h)$, enabling end-to-end functional consistency verification. 

Table~\ref{tab:dataset_quality} (b) reports aggregate statistics of the constructed dataset. The final dataset demonstrates strong structural alignment between $TB_c$ and $TB_h$. Dynamic consistency is enforced as a hard filtering condition, and any testbench pair violating runtime input equivalence is discarded. Therefore, the dynamic consistency rate reaches 1.000.

\input{mic/compare}
\subsection{Ablation Study}
\label{sec:ablation}


\subsubsection{Component Analysis}
The proposed framework relies on two core synergistic modules: the Knowledge Graph for providing epistemic verification priors, and the closed-loop Agent for execution-based reactive refinement. To isolate and quantify their respective impacts, we configure four experimental settings, as shown in Table~\ref{tab:method_ablation}:

\begin{enumerate}[label=\textit{(\roman*)}]
    \item Zero-Shot: The LLM generates the testbench pair $(TB_c, TB_h)$ directly from $(P_c, P_h)$ without knowledge retrieval or iterative feedback.
    \item KG-Only: The LLM is augmented with $\mathcal{G}_{cov}$ and $\mathcal{G}_{sem}$, but no closed-loop refinement is applied.
    \item Agent-Only: The system performs iterative refinement via feedback (up to 3 iterations), without KG-based priors.
    \item Ours (KG + Agent): The complete framework integrating both KG-guided reasoning and Agent-based refinement.
\end{enumerate}

The \textit{Zero-Shot} baseline already achieves relatively high Syntax correctness and reasonable coverage, but its functional consistency remains limited at 0.5771, with a Dynamic Rate of 0.6729. This indicates that although modern LLMs can produce syntactically valid and partially executable testbenches, they struggle to ensure deeper semantic alignment and robust behavioral correctness without external guidance.

Introducing KG brings consistent improvements across all major metrics. 
These improvements suggest that the structured priors encoded in $\mathcal{G}_{sem}$ and $\mathcal{G}_{cov}$ effectively constrain the generation space, reducing semantic mismatches and improving initial testbench quality.

The \textit{Agent-Only} configuration shifts the performance profile toward execution quality. It achieves perfect Syntax correctness and significantly boosts Dynamic Rate to 0.8692. This demonstrates that iterative feedback is highly effective for exploring execution paths and repairing behavioral errors.

The complete framework achieves the best overall performance. Compared to Agent-Only, it shows clear gains in structural alignment, particularly in CFG-based consistency. This confirms that KG provides a more accurate starting point, while the Agent refines residual errors through execution feedback. The combination effectively balances prior-guided generation and runtime correction, leading to consistently superior results.

\subsubsection{Model Generalization Analysis}

Under the \textit{Zero-Shot} setting, model capacity plays a dominant role. 
After applying the proposed framework, all models show substantial and consistent improvements.
A key observation from Table~\ref{tab:method_ablation} is that the performance gap between models is noticeably reduced after introducing the framework. The structured priors and feedback-driven refinement process transform the task from open-ended generation into a more controlled and decomposable reasoning problem. As a result, weaker models benefit disproportionately, while stronger models achieve incremental but consistent gains.

\subsection{Comparison with Baseline Methods}
\label{sec:comparison}

We compare our framework with representative baseline paradigms. The results are summarized in Table~\ref{tab:baseline_comparison}. The baseline methods exhibit a clear performance progression from AST-based retrieval to RAG and iterative Agent frameworks. AST-based and RAG-based methods provide reasonable syntax correctness and coverage, but their functional consistency remains limited due to insufficient semantic alignment. Incorporating execution feedback through iterative Agents substantially improves coverage and dynamic consistency, demonstrating the importance of closed-loop refinement for verification tasks. However, relying solely on iterative optimization is still insufficient to guarantee strong structural and behavioral alignment between generated testbenches.

In contrast, the proposed framework consistently outperforms all baselines across every metric. It achieves perfect Syntax correctness, the highest coverage, and the strongest functional consistency, with a Dynamic Rate of 0.9533. The improvement over the iterative Agent baseline is particularly notable, reflecting the complementary role of the Knowledge Graph in providing precise initialization that reduces the burden on subsequent refinement.

%% file: mic/ablation1.tex
\begin{table*}[t]
\centering
\caption{Method Ablation Study and Model Ablation Study}
\label{tab:method_ablation}
\begin{adjustbox}{width=\textwidth}
\begin{tabular}{p{3.3cm}|c|cccc|cccc|c}
\toprule
 & \multicolumn{1}{c}{\textbf{Syntax}} 
 & \multicolumn{4}{|c|}{\textbf{Coverage}} 
 & \multicolumn{5}{c}{\textbf{Functional Consistency}} \\
\cmidrule(lr){2-2} \cmidrule(lr){3-6} \cmidrule(lr){7-11}
\textbf{Method Ablation} 
& TB & Statement & Branch & Call & Avg. & Input & CFG & DDG & Avg. & Dynamic \\
\midrule
GPT-5-mini (zero-shot) 
& 0.9533 & 0.8877 & 0.9203 & 0.8573 & 0.8884 & 0.6268 & 0.5459 & 0.5588 & 0.5771 & 0.6729 \\

GPT-5-mini + KG 
& 0.9626 & 0.9152 & 0.9387 & 0.8841 & 0.9127 & 0.6645 & 0.6102 & 0.6214 & 0.6320 & 0.7570 \\

GPT-5-mini + Agent 
& \textbf{1.000} & 0.9566 & 0.9756 & 0.9370 & 0.9564 & 0.7138 & 0.7426 & 0.7343 & 0.7302 & 0.8692 \\
\midrule
GPT-5-mini + KG + Agent
& \textbf{1.000} & \textbf{0.9799} & \textbf{0.9820} & \textbf{0.9789} & \textbf{0.9803} & \textbf{0.7421} & \textbf{0.8035} & \textbf{0.7402} &\textbf{ 0.7619} & \textbf{0.9439} \\

\bottomrule
\end{tabular}
\end{adjustbox}

\vspace{0.4em}

\begin{adjustbox}{width=\textwidth}
\begin{tabular}{p{3.3cm}|c|cccc|cccc|c}
\toprule
 & \multicolumn{1}{c}{\textbf{Syntax}} 
 & \multicolumn{4}{|c}{\textbf{Coverage}} 
 & \multicolumn{5}{|c}{\textbf{Functional Consistency}} \\
\cmidrule(lr){2-2} \cmidrule(lr){3-6} \cmidrule(lr){7-11}
\textbf{Model Ablation} 
& TB 
& Statement & Branch & Call & Avg. & Input & CFG & DDG & Avg. & Dynamic \\
\midrule
GLM-5 (zero-shot) 
& 0.7103 & 0.6579 & 0.6591 &  0.6142 & 0.6437 & 0.3486 & 0.3765 & 0.3349 & 0.3534 & 0.4206 \\

GLM-5 + KG + Agent 
& 0.9159 & 0.8183 & 0.8246 & 0.8028 & 0.8153 & 0.4093 & 0.4339 & 0.4500 & 0.4310 & 0.6168 \\

\midrule

GPT-5-mini (zero-shot) 
& 0.9533 & 0.8877 & 0.9203 & 0.8573 & 0.8884 & 0.6268 & 0.5459 & 0.5588 & 0.5771 & 0.6729 \\

GPT-5-mini + KG + Agent 
& \textbf{1.000} & \textbf{0.9799} & 0.9820 & 0.9789 & 0.9803 & 0.7421 & \textbf{0.8035}  & \textbf{0.7402} & \textbf{0.7619} & 0.9439 \\

\midrule

Gemini-3-pro (zero-shot) 
& \textbf{1.000} & 0.9342 & 0.9423 & 0.9171 & 0.9312 & 0.7467 & 0.7273 & 0.7009 & 0.7250 & 0.9065 \\

Gemini-3-pro + KG + Agent 
& \textbf{1.000} & 0.9793 & \textbf{0.9865} & \textbf{0.9818} & \textbf{0.9826} & \textbf{0.7611} & 0.7540 & 0.7299 & 0.7483 & \textbf{0.9533} \\

\bottomrule
\end{tabular}
\end{adjustbox}

\end{table*}

%% file: mic/datascore.tex
\begin{table}[t]
\centering
\caption{Dataset Comparison and Quality Statistics}
\label{tab:dataset_quality}
\setlength{\tabcolsep}{6pt}
\renewcommand{\arraystretch}{1.2}
\textbf{(a) Structural Comparison with Existing Datasets}
\begin{tabular}{l|p{1cm}p{1cm}p{1cm}p{1.5cm}}
\toprule
\textbf{Dataset} & \textbf{C} & \textbf{HLS-C} & \textbf{C-TB} & \textbf{HLS-C-TB} \\
\midrule
HLSDataset     & $\times$   & \checkmark & $\times$ & \checkmark \\
HLS-Eval        & $\times$   & \checkmark & $\times$ & \checkmark \\
HLSTrans        & \checkmark & \checkmark & $\times$ & $\times$ \\
HLSPilot        & \checkmark & \checkmark & $\times$ & $\times$ \\
Ours            & \checkmark & \checkmark & \checkmark & \checkmark \\
\bottomrule
\end{tabular}

\vspace{0.5em}

\textbf{(b) Dataset Quality Metrics}
\begin{tabular}{ccccc}
\toprule
\textbf{\#Samples} & \textbf{Syntax} & \textbf{Coverage} & \textbf{Static Score} & \textbf{Dynamic} \\
\midrule
1597 & 1.000 & 0.9808 & 0.7998 & 1.000 \\
\bottomrule
\end{tabular}

\end{table}

%% file: mic/compare.tex
\begin{table*}[t]
\centering
\caption{Comparison with Representative Baseline Paradigms}
\label{tab:baseline_comparison}
\renewcommand{\arraystretch}{1.15}
\begin{adjustbox}{width=\textwidth}
\begin{tabular}{p{3.3cm}|c|cccc|cccc|c}
\toprule
 & \multicolumn{1}{c}{\textbf{Syntax}} 
 & \multicolumn{4}{|c|}{\textbf{Coverage}} 
 & \multicolumn{5}{c}{\textbf{Functional Consistency}} \\
\cmidrule(lr){2-2} \cmidrule(lr){3-6} \cmidrule(lr){7-11}
\textbf{Method} 
& TB 
& Statement & Branch & Call & Avg. 
& Input & CFG & DDG & Avg. 
& Dynamic \\
\midrule

AST-based KG 
& 0.9720 
& 0.9186 & 0.9321 & 0.9045 & 0.9184 
& 0.6823 & 0.7125 & 0.6339 & 0.6762 
& 0.7383 \\

RAG-based 
& 0.9626
& 0.9141 & 0.9438 & 0.9102 & 0.9227 
& 0.7012 & 0.7064 & 0.6675 & 0.6917 
& 0.7664 \\

Iteration-based Agent 
& 0.9945 
& 0.9382 & 0.9695 & 0.9217 & 0.9431 
& 0.7084 & 0.7352 & 0.7216 & 0.7217 
& 0.8505 \\

\midrule

\textbf{Ours (Gemini-3-pro) }
& \textbf{1.000} & \textbf{0.9793} & \textbf{0.9865} & \textbf{0.9818} & \textbf{0.9826} & \textbf{0.7611} & \textbf{0.7540} & \textbf{0.7299} & \textbf{0.7483} & \textbf{0.9533} \\

\bottomrule
\end{tabular}
\end{adjustbox}
\end{table*}

%% file: sec/conclusion.tex
\section{Conclusion}
\label{sec:conclusion}
This paper presented a knowledge-augmented, agent-driven shift-left verification framework for automated functional consistency checking between original C specifications and HLS-oriented C implementations prior to synthesis. The proposed framework integrates a Dual-Tier Consistency Checking mechanism, symbolic execution and coverage-driven refinement for high-coverage stimulus generation, an autonomous verification Agent for closed-loop orchestration, and a heterogeneous HLS Verification Knowledge Graph that provides structural and semantic reasoning priors for testbench construction. Experimental results on the HLSTrans benchmark demonstrate that the proposed method consistently outperforms representative AST-based, RAG-based, and iterative Agent-based baselines in syntax correctness, coverage, functional consistency, and dynamic equivalence. Future work will extend the framework to RTL-level equivalence checking, multi-module system verification, and adaptive verification scheduling based on reinforcement learning.